\input{aipcheck}

\documentclass[
    ,final            % use final for the camera ready runs
  ]
  {aipproc}

\layoutstyle{6x9}

\begin{document}
\def\pim{$\pm$} 
\def\be{\begin{equation}}
\def\ee{\end{equation}}
\def\mdot{$\dot{m}$ }
\def\mpc{\,{\rm {Mpc}}}
\def\kpc{\,{\rm {kpc}}}
\def\kms{\,{\rm {km\, s^{-1}}}}
\def\msun{{$\rm M_{\odot}$}}
\def\Gyr{{\,\rm Gyr}}
\def\erg{{\rm erg}}
\def\sr{{\rm sr}}
\def\hz{{\rm Hz}}
\def\cm{{\rm cm}}
\def\sec{{\rm s}}
\def\eV{{\rm \ eV}}
\def\ledd{$L_{Edd}$~}
\def\mic{$\mu$ }
\def\ang{\AA }  % Angstrom
\def\cm2{cm$^2$ }
\def\se1{s$^{-1}$ }

\def\arcmin{\hbox{$^\prime$}}
\def\arcsec{\hbox{$^{\prime\prime}$}}
\def\degree{$^{\circ}$} 
\def\mic{$\mu$ }
\def\ang{\AA }  % Angstrom
\def\cm2{cm$^2$ }
\def\se1{s$^{-1}$ }

\def\gtsima{$\; \buildrel > \over \sim \;$}
\def\ltsima{$\; \buildrel < \over \sim \;$}
\def\prosima{$\; \buildrel \propto \over \sim \;$}
\def\gsim{\lower.5ex\hbox{\gtsima}}
\def\lsim{\lower.5ex\hbox{\ltsima}}
\def\simgt{\lower.5ex\hbox{\gtsima}}
\def\simlt{\lower.5ex\hbox{\ltsima}}
\def\simpr{\lower.5ex\hbox{\prosima}}
\def\la{\lsim}
\def\ga{\gsim}
\def\Lsun{\rm L_{\odot}}
\def\Rsun{\rm R_{\odot}} 
\def\sr{V404~Cyg~}
\def\gx{GX~339$-$4~}

\def\ie{{\frenchspacing\it i.e. }}
\def\eg{{\frenchspacing\it e.g. }}
\def\etal{{~et al.~}}

\title{Black hole X-ray binary jets}


\author{Elena Gallo}{
  address={Astronomical Institute `Anton Pannekoek', University of Amsterdam,
Kruislaan 403, 1098 SJ, Amsterdam, the Netherlands}
}

\author{Rob Fender}{
  address={School of Physics and Astronomy,  
University of Southampton,
Hampshire SO17 1BJ Southampton, 
United Kingdom}
}

\author{Christian Kaiser}{
  address={School of Physics and Astronomy,  
University of Southampton,
Hampshire SO17 1BJ Southampton, 
United Kingdom}
% altaddress={<author1 address>} % additional visiting address
}

\begin{abstract}
Relativistic jets powered by stellar mass black holes in X-ray binaries appear
to come in two types: steady outflows associated with hard X-ray states and
large scale discrete ejections associated with transient outbursts.  We show
that the broadband radio spectrum of a `quiescent' stellar mass black hole
closely resembles that of canonical hard state sources emitting at four orders
of magnitude higher X-ray levels, suggesting that a relativistic outflow is
being formed down to at least a few $10^{-6}$ times the Eddington X-ray
luminosity. We further report on the discovery of a low surface brightness
radio nebula around the stellar black hole in Cyg X-1, and discuss how it can
be used as an effective calorimeter for the jet kinetic power
\end{abstract}

\maketitle

%%%%%%%%%%%%%%%%%%%%%%%%%%%%%%%%%%%%%%%%%%%%
%% MAINMATTER
%%%%%%%%%%%%%%%%%%%%%%%%%%%%%%%%%%%%%%%%%%%%

\section{Astrophysical relevance}

The production of jets, collimated bipolar outflows with relativistic
velocities, appears to be a common consequence of accretion of material onto
black holes on all mass scales. However, despite decades of study, we still
lack a comprehensive theory that might account for the mechanism(s) of jet
production, acceleration and collimation, and address the issue of the
coupling between the outflow of matter and the accretion flow
(e.g. \cite{bl01}). The advantage of studying relativistic jets powered by
stellar mass objects is simply given by their rapid variability: as the
physical timescales associated with the jet formation are thought to be set by
the accretor's size, and hence mass, then by observing stellar mass black
holes in Galactic X-ray binary systems (BHXBs) on timescales of days to
decades we are probing the time-variable jet:accretion coupling on timescales
of tens of thousands to millions of years or more for supermassive black holes
at the centres of active galactic nuclei (AGN). Even though such jets are
expected to be highly radiatively inefficient -- being adiabatic expansion the
dominant cooling process -- in the recent years it has become apparent that
they may carry away a significant (if not the dominant) fraction of the
liberated accretion power in low luminosity systems, possibly acting as a
major source of energy and entropy for the interstellar medium.
\section{Radio states of black hole X-ray binaries}

Historically, the key observational aspect of X-ray binary jets lies in their
radio emission (\cite{hh95}, \cite{mirabelrodriguez99}); in BHXBs, different
radio properties are associated with different X-ray spectral states (see
\cite{fender04}, \cite{mcr} and Remillard, in these Proceedings, 
for recent reviews). This is illustrated schematically in Figure 1, which
shows the spectral energy distribution, from radio to $\gamma$-ray
wavelengths, of the (prototypical) stellar mass black hole in Cygnus X-1 over
different accretion regimes.
\begin{figure}
 \includegraphics[height=.35\textheight]{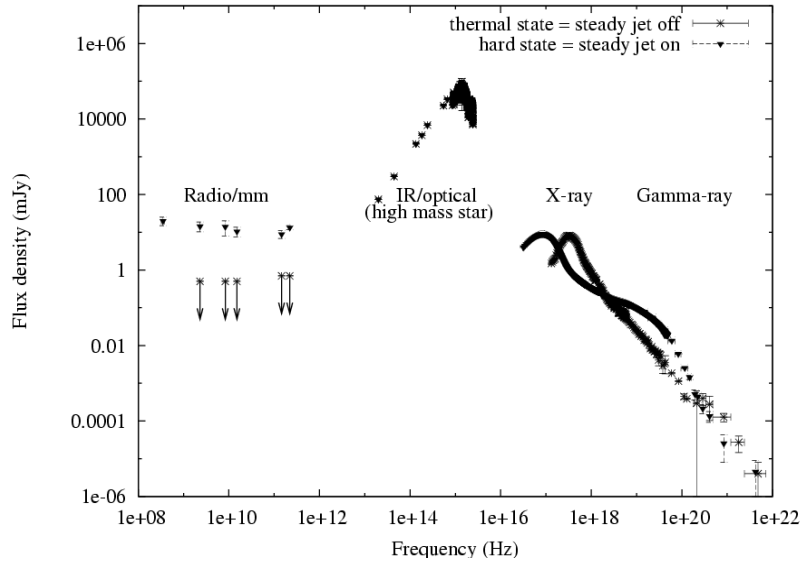}
\caption{Spectral energy distribution of the prototypical 10
$M_{\odot}$ BH in the high mass X-ray binary Cygnus X-1.  When the X-ray
spectrum (above $\sim 10^{18}$ Hz) is dominated by a hard power-law component
(triangle-points), the system is persistently detected in the radio band. The
radio-mm spectrum is flat, due to an inhomogeneous, partially self-absorbed
steady jet resolved on milliarcsec-scales \cite{stirling01}. Above a critical
X-ray luminosity, the disc contribution becomes dominant (in units of $\nu
F_{\nu}$), while the hard X-ray power law softens (star-points). In this
`thermal dominant' state the radio emission is quenched by a factor up to
about 50 with respect to the hard state. Adapted from \cite{tigelaar}.}
\end{figure}
%\subsection{Steady jets}

While accreting gas at relatively low rates (below a critical X-ray luminosity
of a few per cent of the Eddington X-ray luminosity, $L_{\rm Edd}$), BHXB
systems emit the bulk of their radiation in form of a hard X-ray power law
component that cuts off at a few hundreds of keV, traditionally interpreted as
due to Comptonization of disc photons in a rarefied electron/positron plasma
(\cite{st}, \cite{tit}). In terms of radio properties this {\it hard} X-ray
state is {\it radio active}, associated with persistent emission and a flat or
slightly inverted spectrum which extends up to near-IR (possibly optical)
frequencies, thought to be synchrotron in origin (\cite{fender01} and
ref. therein).  The outflow nature of the synchrotron-emitting (and thus
relativistic) plasma is inferred by brightness temperature arguments, leading
to minimum linear sizes for the emitting region that often exceed the typical
orbital separations, and thus making it unconfinable by any known component of
the binary. This, together with the fact that BHXBs in the hard state display
persistent radio emission despite being inevitably subject to expansion
losses, imply the presence of a continuously replenished relativistic plasma
that is flowing out of the system. The flat radio spectrum would arise in a
{\it steady} partially self-absorbed conical jet, becoming progressively more
transparent at lower frequencies as the matter travels away from the launching
site (\cite{bk79},
\cite{hj}). 
%Observed time delays between different frequencies
%(e.g. in GRS 1915+105: \cite{pf97}, \cite{mirabeletal98}) rule out models in
%which the flat radio spectrum would be due to to optically thin synchrotron
%emission from a very hard energy distribution of electrons (\cite{wang}). 
The collimated nature of these outflows is less certain, as it requires direct
imaging to be proven. Milliarcsec-resolution observations of Cyg X-1 in the
hard X-ray state have confirmed the jet interpretation of the flat radio-mm
spectrum in this system, imaging an extended structure extending to about 15
milliarcsec ($\sim$30 A.U. at 2 kpc), and with an opening angle of less then
2\degree (\cite{stirling01}).

%The presence of a jet can also be inferred by its long-term action on the
%local interstellar medium, as in the case of the hard state BHXBs
%1E1740.7$-$2942 and GRS~1758$-$258, both associated with arcmin-scale radio
%lobes (\cite{mirabeletal92}, \cite{rodriguez92}).  
Further indications for the
existence of collimated outflows in the hard state of BHXBs come from the
stability in the orientation of the electric vector in the radio polarization
maps of GX~339$-$4 over a two year period (\cite{corbel00}). This constant
position angle, being the same as the sky position angle of the large-scale,
optically thin radio jet powered by GX 339$-$4 after its 2002 outburst
(\cite{gallo04}), clearly indicates a favoured ejection axis in the system.
Finally, the milliarcsec scale jet of the (somewhat peculiar) BH candidate GRS
1915+105 (\cite{dhawan00}; \cite{fuchs03}) in the hard state
supports the association of hard X-ray states of BHXBs with steady, partially
self-absorbed jets.

It is worth mentioning that some authors propose a jet interpretation (rather
than a Comptonizing `corona') for the X-ray power law which dominates the
spectrum of BHXBs in the hard/quiescent (see next Section) state (\cite{mff},
\cite{markoffnowak}). In this model, depending on the location of the 
frequency above which the jet synchrotron emission becomes optically thin to
self-absorption and
the distribution of the emitting particles, a significant fraction -- if not the
whole -- of the hard X-ray photons would be produced in the inner regions of the
steady jet, by means of optically thin synchrotron and synchrotron
self-Compton emission. \\
%\subsection{Transient jets}

Above a few per cent of $L_{\rm Edd}$, BHXBs enter the so called {\it thermal
dominant} X-ray state (starred points in Figure 1), during which the power
output is dominated by a thermal component with typical temperatures of about
1 keV, interpreted as the clear signature of a geometrically thin optically
thick accretion disc (\cite{ss}) extending very close to the central hole. No
core radio emission is detected while in the thermal dominant state: the radio
fluxes are {\it quenched} by a factor up to about 50 with respect to the hard
X-ray state (\cite{fender99}, \cite{corbel01}), probably corresponding to the
physical disappearance of the steady jet.  This has been taken as a strong
arguments in favour of magneto-hydro-dynamic jet formation in geometrically
thick accretion flows (\cite{meier01}).\\

Additionally, we observe a second variety of radio jets powered by BHXBs:
hard-to-thermal X-ray state transitions appear to be associated with arcsec
scale (thousands of A.U.) synchrotron-emitting plasmons moving away from the
binary core with highly relativistic velocities (\cite{mirabelrodriguez99},
\cite{fender04} and ref. therein). Unlike milliarcsec scale steady jets, such
discrete ejection events display optically thin synchrotron spectra and
rapidly decaying fluxes. We shall refer to them as {\it transient} jets.

\section{Relativistic outflows in `quiescence'}
What are the required conditions for a steady jet to exist? We wonder
especially whether the steady jet survives in the very low luminosity,
\emph{quiescent} X-ray state (with $L_{\rm X}\simlt 10^{33.5}$ erg sec$^{-1}$,
i.e. below a few $10^{-5} L_{\rm Edd}$). In such a regime, very
few systems have been detected in the radio band, mainly because of
sensitivity limitations on the existing telescopes.  Among them, the faintest
is V404 Cygni, hosting a 12
\msun~ black hole (\cite{shahbaz94}) and emitting in the X-rays at a few 
$10^{-6} L_{\rm Edd}$ (\cite{kong02}). Given the quite large degree of
uncertainty about the overall structure of the accretion flow in quiescence
(see \cite{mcr} and ref. therein), it has even been speculated that the total
power output of a quiescent BH could be dominated by a radiatively inefficient
outflow (\cite{gfp}, \cite{fgj}) rather than by the local dissipation of
gravitational energy in the accretion flow. It is therefore of primary
importance to establish the nature of radio emission from quiescent BHXBs.

Radio observations (using the Westerbork Synthesis Radio Telescope, WSRT) of
this system, performed on 2002 December 29 (MJD 52637.3) at four frequencies,
over the interval 1.4$-$8.4 GHz, have provided us with the
\emph{first} broadband radio spectrum of a quiescent stellar mass black
hole. We measured a mean flux density of 0.4 mJy, consistent with that
reported by \cite{hj00}, and a flat/inverted spectral index $\alpha=
0.09\pm0.19$ (such that $\rm S_{\nu}\propto
\nu^{\alpha}$). WSRT observations performed one year earlier, at 4.9 and 8.4
GHz, resulted in a mean flux density of 0.5 mJy, confirming the relatively
stable level of radio emission from \sr on a year time-scale; even though the
spectral index was not well constrained at that time, the measured value was
consistent with the later one.

Synchrotron emission from a partially self-absorbed relativistic outflow of
plasma seems to be the most likely explanation for the flat radio spectrum.
Optically thin free-free emission as an alternative is ruled out, on the basis
that far too high mass loss rates would be required in order to sustain the
observed radio flux: even taking into account geometrical effects, such as
outflow collimation and/or clumpiness, the mass loss rates can not be lower
than $10^{-3}$ times the Eddington rate (assuming a 10 per cent efficiency in
converting mass into light), \ie still far too high for a sub-$10^{-5}$
Eddington BH to produce any observable radio emission (see \cite{gfh} for
details).

The collimated nature of this outflow remains to be proven; based on
brightness temperature arguments and the 5.5-hour time-scale variability
detected at 4.9 GHz, we conclude that the angular extent of the radio source
is constrained between 0.01 at 1.4 GHz and 10 mas at 4.9 GHz (at a distance of
4 kpc;
\cite{peter}).  These arguments led us to suggest that a
relativistic jet is being formed in the quiescent state of V404 Cyg, and
probably in BHXBs between a few $10^{-6}$ and a few per cent of $L_{\rm Edd}$
(were the collimated jet is actually resolved), strengthening the notion of
`quiescence' as a low luminosity version of the canonical hard X-ray state
(\cite{gfh}).

\begin{figure}
 \includegraphics[height=.4\textheight]{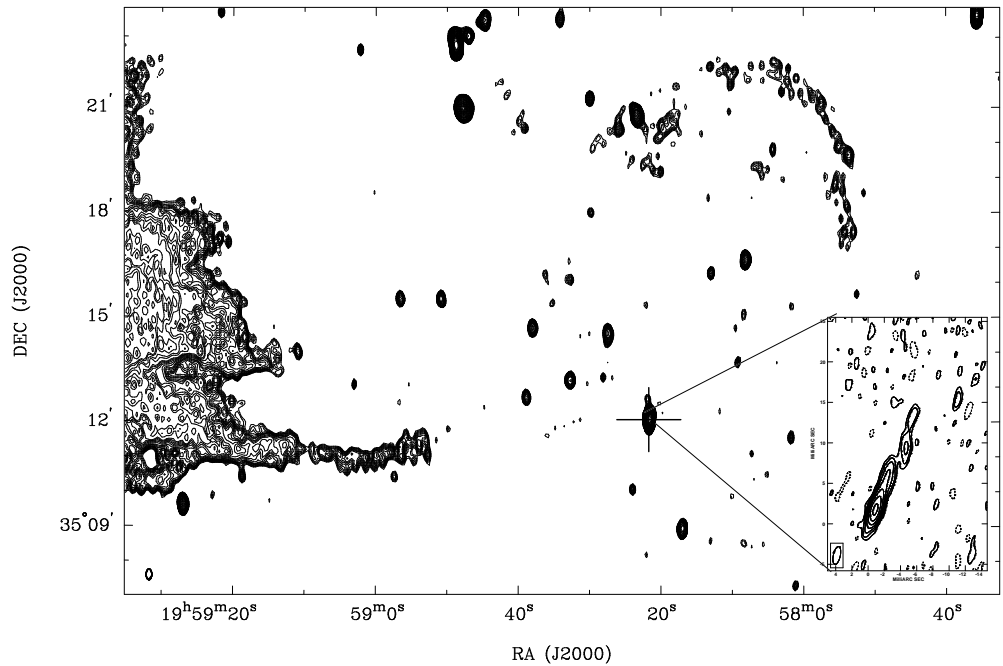}
\caption{Westerbork Synthesis Radio Telescope 1.4 GHz of the field of view of
the BHXB and Galactic jet source Cyg X-1: the arcmin scale, semi-ring-like
structure northeast of the binary core (marked by a cross) seems to draw an
edge between the bright HII region west of Cyg X-1 and the direction of the
Cyg X-1 milliarcsec scale jet, shown in the inset (VLBA map from
\cite{stirling01}); the average monochromatic flux of the ring is of about 0.08
mJy/beam. We interpret this structure as the result of a strong shock that
develops at the location where the collimated jet impacts on the ambient ISM
(Gallo, Fender, Kaiser \& Russell, in preparation).}
\end{figure}

\section{A unified picture}
The question remains whether the steady and transient jets of
BHXBs have a different origin or are somewhat different
manifestations of the same phenomenon. \cite{fbg} have
addressed this issue, showing that: i) the power content of the steady and
transient jets are consistent with a monotonically increasing function of
$L_{X}$; ii) the measured bulk Lorentz factors of the transient jets are
systematically higher than those inferred for the steady jets. Based upon
these arguments, a unified model for the jet/accretion coupling
in BHXBs has been put forward. The key idea is that, as the disc inner
boundary moves closer to the hole (hard-to-thermal state transition), the
escape velocity from the inner regions increases.  As a consequence, the
steady jet bulk Lorentz factor rises sharply, causing the propagation of an
{\em internal shock} through the slower-moving outflow in front of it.
Eventually, the result of this shock is what we observe as a post-outburst,
optically thin radio plasmon. For a thorough description of the model, we
refer the reader to Belloni, in these Proceedings.

\section{Discovery of a jet-powered radio nebula in Cyg X-1}
The importance of BHXB jets for the overall energetics and dynamics of the
accretion process, and furthermore as a potentially major source of energy
input into the galactic interstellar medium (ISM), has yet to be well
quantified. For both the steady and transient jets we are forced to make
assumptions about the spectral extent (as the jet high frequency emission is
generally blocked by the accretion disc or the companion star) and radiative
efficiency, basing our estimates on, for example, assumptions of equipartition
for which there is little a priori justification. Alternatively, we can
constrain the jets' power content by looking at their (gradual or abrupt)
interaction with the surrounding medium. As in AGN, the total energy
associated with radio lobes and termination shocks was, and to a certain
extent remains, the safest way to estimate the jet {\it power$\times$lifetime}
product (\cite{burbidge59}).  In the case of Galactic stellar mass BHs,
arcmin-scale radio lobes are associated with two hard state sources in the
galactic centre, 1E 1740.9-2942 and GRS 1758-258 (\cite{mirabeletal92},
\cite{rodriguez92}).  \cite{chaty01} identified two IRAS sources with flat
spectrum symmetric about GRS1915+105 (see also \cite{rm98}) but argued that a
possible association with the arcsec scale jets in these system seemed
inconclusive.
\cite{kaiser04} suggest instead that the two IRAS regions would be the actual 
jets' impact sites; applying a fluid dynamical model developed for
AGN jets (\cite{ka97}) to this stellar mass system, they conclude
that the time-averaged energy transport rate in the jet may be as low as
$10^{36}$ erg 
\se1 (i.e. at least one order of magnitude than what inferred for the discrete
ejecta;\cite{fender99}). If correct, this association would place GRS 1915+105
at the same distance of the two IRAS sources, i.e. at 6.5 kpc (rather than the
12 kpc estimated by \cite{greiner}), casting doubts even on its `superluminal'
nature. Finally, we have recently witnessed the dynamic formation of
arcmin-scale decelerating radio {\it and} X-ray lobes, following an outburst of
the transient BHXB XTE J1550-564 (\cite{corbel02}).  These results suggest
that radio lobe formation is a common occurrence which might be associated
with many more sources and has not been found to date due to low
signal-to-noise.

Such considerations led us to look for extended radio nebulae around the more
promising (i.e. radio-loud) binary systems powering jets with the
WSRT. Observations performed in May 2003 at 1.4 GHz resulted in the {\em
discovery of an arcmin scale semi-ring-like radio nebula around the Galactic
jet source Cyg X-1}, presented in Figure 2 (Gallo, Fender, Kaiser \& Russell,
in preparation). The structure appears to be perfectly aligned with the
collimated jet resolved on milliarcsec scale. We note that previous attempts
to look for lobes around Cyg X-1 (\cite{martietal96}) resulted in several
`interesting structures' at 1.4 GHz, whose association with Cyg X-1 could not
be proven yet. Interestingly, Mart\'\i~ and collaborators (1996) already
talked of a `suggestive shell appearance of the structures' around Cyg X-1,
which led them to investigate the possibility of a weak supernova remnant
interpretation, eventually excluded due to too low surface brightness.

\subsection{Modelling the ring of Cyg X-1: jet power and lifetime}

Following the self-similar model developed by \cite{ka97} for extragalactic
jet-cocoon systems, we have interpreted the semi-ring radio structure of Cyg
X-1 as the result of a strong shock that develops at the location where the
collimated jet (resolved on milliarcsec scales) impacts on the ambient
ISM. The jet particles inflate a radio lobe which is over-pressured with
respect to the surrounding medium, thus the lobe expands sideways forming the
observed bow shock that emits bremsstrahlung radiation -- {\em hypothesis
which needs to be confirmed through approved optical and deep 90~cm 
observations of this field}. The very pressure in the cocoon is responsible
for keeping the jet confined.  Following their formalism, we assume:
\begin{enumerate}
\item
that the jet and the shocked ISM are in pressure balance;
\item
that the rate of energy input is constant and given by the average jet power,
$Q_0$;
\item
that the rate of mass transport along the jet is constant and
supplied 
by the bulk kinetic energy of the jet;
%, such that $\dot M_0 = Q_0/[(\Gamma_j -
%1) c^2]$, where $\Gamma_j$ is the (constant) bulk Lorentz factor of the jet; 
\item
that the shock expands into an atmosphere of constant mass density
$\rho_0$.
\end{enumerate}
This model implicitly requires the bow shock to be self-similar and a roughly
constant jet direction (which seems to be the case here, as the measured
proper motion of Cyg X-1 rules out large velocity kicks;
\cite{mirabel03}).

Knowing the ring monochromatic luminosity, we are able to derive the density
of the ionized particles in the ring from the expression of the bremsstrahlung
emissivity $\epsilon_{\nu}$\footnote{$\epsilon_{\nu}=6.8\times 10^{-38}~
g(\nu,T) ~T^{(-1/2)}~ n_e^2~ {\rm exp} (h\nu/k_BT)~~~\rm
erg~cm^{-3}~sec^{-1}~Hz^{-1}  $} for a pure hydrogen gas emitting at a typical
temperature of $T\simeq 10^4$ K (below which the ionization fraction becomes
negligible, and above which the cooling time becomes too short). 
The measured luminosity density, $L_{1.4~\rm GHz}\simeq 4.8\times 10^{17}$ erg
\se1 \hz~(estimated assuming a distance of 2 kpc to Cyg X-1), 
equals the product ($\epsilon_{\nu}\times V$), where the source volume $V$ is
given by the beam area times the measured ring thickness: $V
\simeq 2.0 \times 10^{53}$ cm$^3$. For $T\simeq 10^4$ K and a Gaunt factor 
$g\simeq 6$, we derive a particle density $n_e$ of the ionized particles of
$\sim 24$  
cm$^{-3}$.  The \emph{total} particle density in the bow shock region will be
a factor $1/x$ higher though, where $x$ is the ionization fraction. At $\sim
10^4$ K, $x=0.019$ (\cite{spitzer}), resulting in a particle density
$n_t$ of $1260$ cm$^{-3}$.\\ 
For a strong shock in a mono-atomic gas, the velocity of the bow shock,
$v_{bow}$, depends on the temperature $T$ of the shocked gas as:
$v_{bow}=\sqrt{ ( 16 k_B / 3m_p) T}$. 

%\end{equation}
For $10^4$ K, $v_{bow}\simeq 2.1\times 10^6$ cm
sec$^{-1}$, justifying the strong shock assumption.  
In \cite{ka97} the length $L$ of the jet within the cocoon grows with the time
in 
such a way that:
$
t = (L/c_1)^{(5/3)}\times(\rho_0 / Q_0)^{(1/3)}$
(where the factor $c_1$ depends on the thermodynamical properties of the jet
material and on the aspect ratio; here $c_1\simeq 1.5$ ). By writing the time
derivative of the above equation, we obtain 
%\begin{equation}
$t=\frac{3}{5}(L / v_{bow})$~,~
%\end{equation}
resulting in a jet lifetime of $\sim$ 0.2 $\times (\rm
{sin}\theta)^{-1}$ Myr, where $\theta$ is the jet angle to the line of
sight. This value has to be compared with the estimated age of the progenitor
of the black hole in Cyg X-1, of a few Myr (\cite{mirabel03}).  For $t\simeq
0.3$ Myr (obtained with $\theta=35$\degree; \cite{orosz}), and adopting a mass
density $\rho_0$ of the un-shocked material that is $\sim$4 times lower than
that of the shocked material, we obtain an average jet power $Q_0$ 
of a few $10^{35}$ erg \se1, which would be a significant fraction of
the measured 0.1-200~keV X-ray power of Cyg X-1 at the peak of the hard X-ray
state (\cite{disalvo01}). The results presented here clearly illustrate that
finding and measuring jet-powered nebulae in stellar mass black holes offers
an alternative and valuable method to address the debated issue of the jet
power content in these systems.

%%%%%%%%%%%%%%%%%%%%%%%%%%%%%%%%%%%%%%%%%%%%
%% SAMPLE TABLE
%%
%% Shows the use of \tablehead and \tablenote
%% macros
%%%%%%%%%%%%%%%%%%%%%%%%%%%%%%%%%%%%%%%%%%%%

%%%%%%%%%%%%%%%%%%%%%%%%%%%%%%%%%%%%%%%%%%%%%%%%
%% BACKMATTER
%%%%%%%%%%%%%%%%%%%%%%%%%%%%%%%%%%%%%%%%%%%%%%%%

\begin{theacknowledgments}
The Westerbork Synthesis Radio Telescope (WSRT) is operated by the ASTRON
(Netherlands Foundation for Research in Astronomy) with support from the
Netherlands Foundation for Scientific Research (NWO).  Raffaella Morganti and
Tom Oosterloo are gratefully acknowledged for their support in the analysis of
the Cyg X-1 data.
%EG wishes to thank the organizers of this meeting for..
\end{theacknowledgments}



\bibliographystyle{aipprocl} % if natbib is missing


%\bibliography{sample}

%  \typeout{******************************************}
%  \typeout{** Please run "bibtex \jobname" to optain}
%  \typeout{** the bibliography and then re-run LaTeX}
%  \typeout{** twice to fix the references!}
%  \typeout{******************************************}

%\end{document}

%%%%%%%%%%%%%%%%%%%%%%%%%%%%%%%%%%%%%%%%%%%
%% The following lines show an example how to produce a bibliography
%% without the help of the BibTeX program. This could be used instead
%% of the above.
%%%%%%%%%%%%%%%%%%%%%%%%%%%%%%%%%%%%%%%%%%%

%\endinput
%%
\end{document}